\documentclass[journal=acsnano,manuscript=article,layout=traditional]{achemso}

\usepackage{chemformula} 
\usepackage[T1]{fontenc} 
\usepackage[colorlinks=true,citecolor=blue,urlcolor=blue,linkcolor=blue,bookmarksopen]{hyperref}

\author{Tommi Isoniemi}
\altaffiliation{Contributed equally to this work as first authors}
\email{t.isoniemi@sheffield.ac.uk}
\author{Paul Bouteyre}
\altaffiliation{Contributed equally to this work as first authors}
\email{p.bouteyre@sheffield.ac.uk}
\author{Xuerong Hu}
\altaffiliation{Contributed equally to this work as first authors}
\author{Fedor Benimetskiy}
\affiliation[University of Sheffield]
{Department of Physics and Astronomy, University of Sheffield, S3 7RH, UK}
\author{Yue Wang}
\affiliation[University of York]
{School of Physics, Engineering and Technology, University of York, YO10 5DD, UK}
\author{Maurice S. Skolnick}
\author{Dmitry N. Krizhanovskii}
\author{Alexander I. Tartakovskii}
\email{a.tartakovskii@sheffield.ac.uk}
\affiliation[University of Sheffield]
{Department of Physics and Astronomy, University of Sheffield, S3 7RH, UK}

\title{Realization of Z$_2$ topological photonic insulators made from multilayer transition metal dichalcogenides}

\abbreviations{TMD, SEM, AFM, EBL, RIE, BIC}

\begin{document}


\begin{abstract}

Monolayers of semiconducting transition metal dichalcogenides (TMDs) have long attracted interest for their intriguing optical and electronic properties. Recently TMDs in their quasi-bulk form have started to show considerable promise for nanophotonics thanks to their high refractive indices, large optical anisotropy, wide transparency windows reaching to the visible, and robust room temperature excitons promising for nonlinear optics. Adherence of TMD layers to any substrate via van der Waals forces is a further key enabler for nanofabrication of sophisticated photonic structures requiring heterointegration.  
Here, we capitalize on these attractive properties and realize topological spin-Hall photonic lattices made of arrays of triangular nanoholes in 50 to 100 nm thick WS$_2$ flakes exfoliated on SiO$_2$/Si substrates. High quality structures are achieved taking advantage of anisotropic dry etching dictated by the crystal axes of WS$_2$. Reflectance measurements at room temperature show a photonic gap opening in the near-infrared in trivial and topological phases. Unidirectional propagation along the domain interface is demonstrated in real space via circularly polarized laser excitation in samples with both zigzag and armchair domain boundaries. Finite-difference time-domain simulations are used to interpret optical spectroscopy results. Our work opens the way for future sophisticated nanophotonic devices based on the layered (van der Waals) materials platform.


\end{abstract}




\section{Introduction}

Over the last two decades, layered crystals, often referred to as van der Waals materials, have attracted tremendous interest due to their unique properties in mono- and few-layer forms.  In particular, semiconducting transition metal dichalcogenides (TMDs) exhibit robust excitons with high oscillator strengths as well as direct bandgaps in monolayers making them attractive for integration in various photonic structures \cite{KinFaiMak2016}. Examples of such integration of TMD mono- and bilayers include realization of exciton-, trion- and dipolar polaritons in dielectric microcavities \cite{liu2015,dufferwiel2015,Datta2022,Louca2023}, lasing in III-V semiconductor nanocavities \cite{Wu2015}, single photon emitters in  monolayers coupled to III-V semiconductor nanoantennas \cite{sortino2021}, and polaritons in spin-Hall topological photonic crystals made from silicon-on-insulator \cite{li2021experimental}.\\


The quasi-bulk counterparts of 2D materials have been much less explored but have recently started attracting considerable attention for their favourable optical properties with a potential for photonic applications (see e.g. \cite{Froch2019,Munkhbat2022devices,zotev2023van}). Similar to monolayers, van der Waals layers of any thickness from few atomic layers up to 100s of nanometers and lateral sizes up to few 100 micrometers can be readily fabricated via mechanical exfoliation \cite{Froch2019,Munkhbat2022devices,zotev2023van}. Thanks to van der Waals forces, the exfoliated flakes can easily adhere to a wide range of substrates without the need for chemical bonding or lattice matching \cite{Froch2019,Munkhbat2022devices,zotev2023van,Liu2019}. By now, there are many demonstrations of standard electron-beam lithography followed by wet or dry etching used to pattern 2D materials leading to high quality structures \cite{munkhbat2020transition,Froch2019,Munkhbat2022devices,zotev2023van,Ling2023}. Furthermore, such patterning can take advantage of etching anisotropy to produce crystallographically exact edges \cite{munkhbat2020transition,zotev2022transition,zotev2023van}.\\

Compared to Si or III-V semiconductors, semiconducting TMDs exhibit higher refractive indices \cite{Munkhbat2022,zotev2023van}, enabling confinement of light to smaller volumes; far larger birefringence values \cite{Munkhbat2022,zotev2023van}, attractive for light polarization control and nonlinear optics; transparency in the visible and near-infrared \cite{Munkhbat2022,zotev2023van}; out-of-plane van der Waals adhesive forces which offer additional post-fabrication tuning \cite{zotev2022transition} and novel approaches to structure fabrication such as vertical layer and structure stacking and twisting \cite{zotev2023van} similar to few-atomic-layer thick van der Waals heterostructures \cite{Geim2013}, which may enable the realization of previously inaccessible photonic structures. Finally, strong room temperature excitonic transitions in most semiconducting van der Waals materials \cite{Munkhbat2022,zotev2023van} open their potential for nonlinear nanophotonic elements.  \\

Having demonstrated the range of favourable material properties offered by van der Waals crystals, in this work, we introduce these materials in the realm of topological photonics. This field emerged following the ideas that were first developed to understand topological phases of matter in the solid state physics starting with the discovery of the integer quantum Hall effect and then, 20 years later, of topological insulators \cite{Hasan2010}. Evolution of the concepts initially applied to electrons led to engineering of artificial magnetic fields (gauge fields) acting on photons and created using specially designed modulated photonic lattices \cite{lu2014topological,khanikaev2017two,barik2018topological,ozawa2019topological,smirnova2020nonlinear,Shalaev2019,parappurath2020direct,mehrabad2020chiral}. Topological photonic devices introduce new functionality in nonlinear and quantum photonic applications, thanks to the unidirectional photonic edge states at their interfaces, which have inherently low scattering losses and backscattering-immune propagation, allowing, for example, scattering-free transport of light around tight bends, and the possibility to form chiral light–matter interfaces \cite{lu2014topological,khanikaev2017two,barik2018topological,ozawa2019topological,smirnova2020nonlinear,Shalaev2019,parappurath2020direct,mehrabad2020chiral}. Photonic analogs of the spin-Hall \cite{wu2015scheme} and valley-Hall \cite{Ma2016} effects have been proposed. In experiment, such topological photonic interfaces have been demonstrated in photonic crystal structures fabricated in standard silicon-on-insulator wafers \cite{Shalaev2019,parappurath2020direct} and suspended GaAs membranes  \cite{barik2018topological,mehrabad2020chiral}.\\

Here, we have designed and fabricated topological spin-Hall lattices in quasi-bulk WS$_2$ exfoliated straight on a SiO$_2$/Si substrate. Our approach resembles the robustness of the silicon-on-insulator platform\cite{Shalaev2019,parappurath2020direct}, with a crucial advantage of a wide range of thicknesses of WS$_2$ crystals available in a single exfoliation step allowing additional degree of freedom, the layer thickness, in designing topological structures for a given target wavelength. Use of a high-refractive-index van der Waals material also allows us to avoid the complication of fabrication of nanophotonic structures in ultrathin suspended membranes necessary for GaAs\cite{barik2018topological,mehrabad2020chiral}.  \\

We have directly observed topological photonic states in the near-infrared spectral range in WS$_2$ spin-Hall lattices with armchair and zigzag shaped domain walls. Topological interface modes are observed both in angle-dependent reflectivity measurements and resonant real-space propagation. Unidirectional propagation lengths as large as 9 \textmu m are recorded, in spite of the use of inherently leaky spin-Hall structures in this initial demonstration. We carry out finite-difference time-domain simulations to reproduce and interpret optical spectroscopy results, both describing the photon bandstructure and real-space propagation. Our work introduces sophisticated nanophotonic devices in the van der Waals materials platform.  \\

\section{Results and Discussion}



\begin{figure*}[!ht] 
\centering
   \includegraphics[width=1\linewidth]{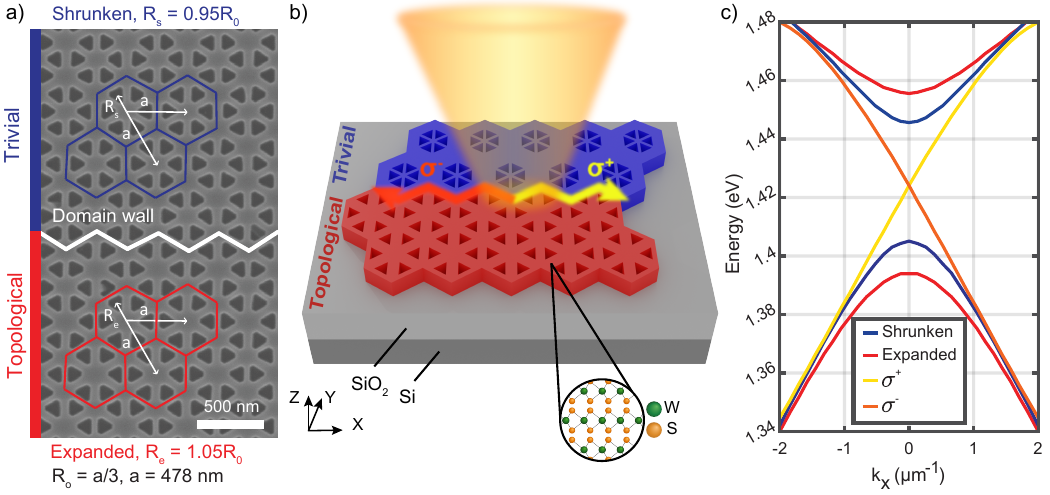}
  \caption{a) Schematic of the hole lattices showing the trivial and topological domains and domain wall overlaid on a scanning electron micrograph of patterned WS$_2$ flake with a thickness 71 nm on SiO$_2$/Si. b) Schematic diagram of the structure and optical excitation at the interface. c) Photonic band structure of the device in a) at $k_y$ = 0 in different domains and at the interface where two modes polarized in $\sigma^+$ and $\sigma^-$ are observed ($x$ is the direction of the interface).}
  \label{fgr:topo}
\end{figure*}

The considered photonic structures, presented in Figure \ref{fgr:topo}a and b, consist of WS$_2$ flakes, used as slab waveguides confined by total internal reflection, patterned as topological lattices. The topological spin-Hall lattice consist of assemblies of 6 triangle-shaped holes patterned on the flake \cite{li2021experimental}, which when arranged in a symmetric orientation, lead to a photonic band structure with a gapless Dirac cone at the $\Gamma$ point. In the shrunken configuration, the lattice is perturbed by moving the triangles inward at each unit cell, which opens the gap with modes of trivial topological nature. In the expanded configuration, the triangles are moved outward, opening the gap with inverted bands compared to the shrunken configuration leading to non-trivial topology (Fig.~\ref{fgr:topo}c). At the interface of the two topologically distinct regions, two states appear within the gap of the band structure, which allow propagating modes.\cite{barik2016two,barik2018topological} Due to the spin-Hall effect, the interface states are accessible for either the counter-clockwise ($\sigma^-$) or clockwise ($\sigma^+$) polarizations, allowing the control of the propagation direction (Fig~\ref{fgr:topo}b). The considered spin-Hall structure is leaky since the states are above the light cone at $\Gamma$ point. This allows direct probing of optical modes using reflectivity measurements.\cite{mehrabad2020chiral} Topological interface states can also be realized in valley-Hall structures, where the states are at K point below the light cone. This allows for better confinement, but separate coupler structures would be required for the detection of the edge states.\cite{he2019silicon} \\ 

\begin{figure*}[!ht] 
    \includegraphics[width=\textwidth]{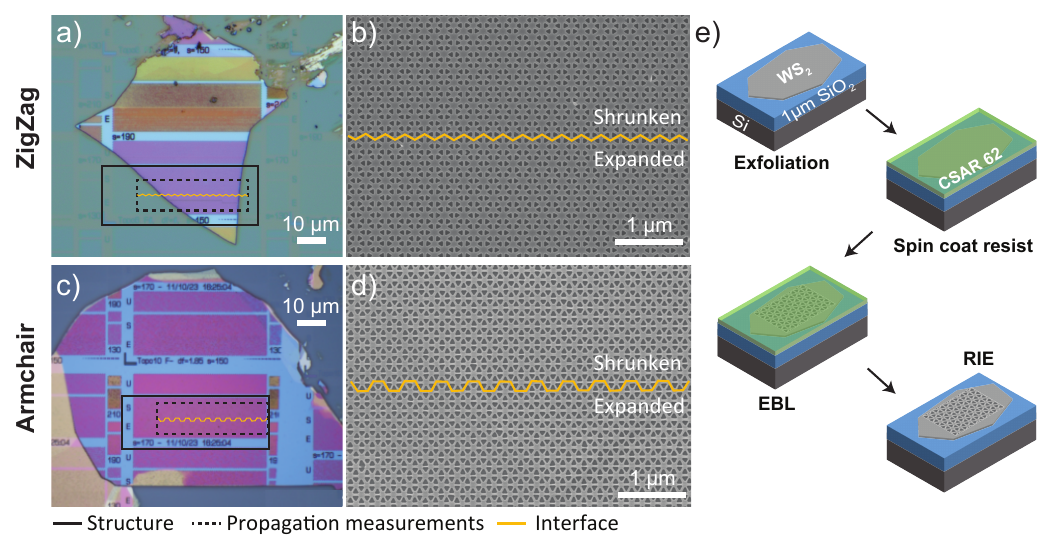}
  \caption{Topological structures with lattice period $a$ = 478 nm etched in a WS$_2$ flake placed on a 1 \textmu m SiO$_2$/Si substrate. a) Optical and b) SEM images of a structure with a zigzag interface (thickness $t$ = 71 nm, triangle side length $s$ = 165 nm, later referred to as the zigzag sample) and c) optical and d) SEM images of a structure with an armchair interface ($t$ = 65 nm, $s$ = 185 nm, armchair sample). e) Schematic depiction of the fabrication procedure.}
  \label{fgr:fab}
\end{figure*}

Two different designs were considered, one with a zigzag domain wall between the shrunken and expanded regions (Figure \ref{fgr:fab}a,b) and one with an armchair domain wall (Figure \ref{fgr:fab}c,d). The slab waveguides were designed to have triangular holes of sizes ranging from 130 to 190 nm, with a period $a$ of 478 nm, in a WS$_2$ layer with thickness $t$ close to 70 nm, situated on top of 1 \textmu m SiO$_2$ on a silicon substrate. For the field to be satisfactorily confined within the waveguide and the propagation to be feasible, a minimum thickness of 50 nm needs to be used for the WS$_2$ layer. 

The topological and trivial lattices were designed to have gaps at around 1.5 eV, set to be significantly lower than the main absorption peak of bulk WS$_2$, associated with its direct transition near 2.0 eV \cite{zotev2023van}. The individual photonic hole lattices are perturbed in terms of expansion ($R_e$ = 1.05$R_0$) and contraction ($R_s$ = 0.95$R_0$) of the individual hexagonal cells as detailed in Fig. \ref{fgr:topo}a. As shown in Figure \ref{fgr:fab}e), the structures were fabricated with the use of electron beam lithography (EBL) and reactive ion etching (RIE) processes on exfoliated \ch{WS_2} flakes (see Materials and Methods). Designs based on the equilateral triangle sides aligned with the \ch{WS_2} crystal axis were patterned with varying doses and triangle sizes on flakes with different thicknesses. In this paper, two structures with zigzag and armchair domain walls of respective triangular side lengths of 165 and 185 nm are considered, as shown in scanning electron microscope (SEM) images in Fig.~\ref{fgr:fab}b,d.\\



\begin{figure*}[!ht] 
\includegraphics[width=\linewidth]{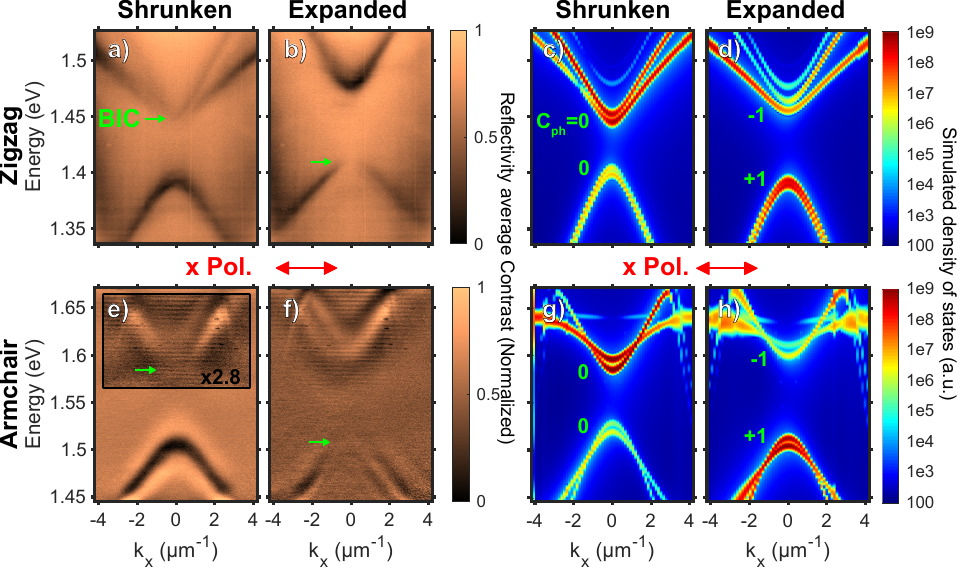}
  \caption{Angle-resolved reflectance contrast data for individual lattices in the zigzag \ch{WS_2} sample (detailed in Fig.~\ref{fgr:fab}) with a) shrunken and b) expanded lattice regions measured with linear polarization along the direction of the interface, $x$. c,d) Photonic band structure simulations of the zigzag sample. e,f) Corresponding reflectance contrast data and g,h) simulations of the armchair sample. The bound states in the continuum (BIC) are highlighted in the measured spectra with green arrows. The spin-Chern numbers $C_{ph}$ of the bands obtained in the simulations are indicated on the plots in green. Note, that simulations report near-field intensities in contrast to the far-field reflectance contrast spectra.}
  \label{fgr:bands}
\end{figure*}
\begin{figure*}[!ht] 
\includegraphics[width=\linewidth]{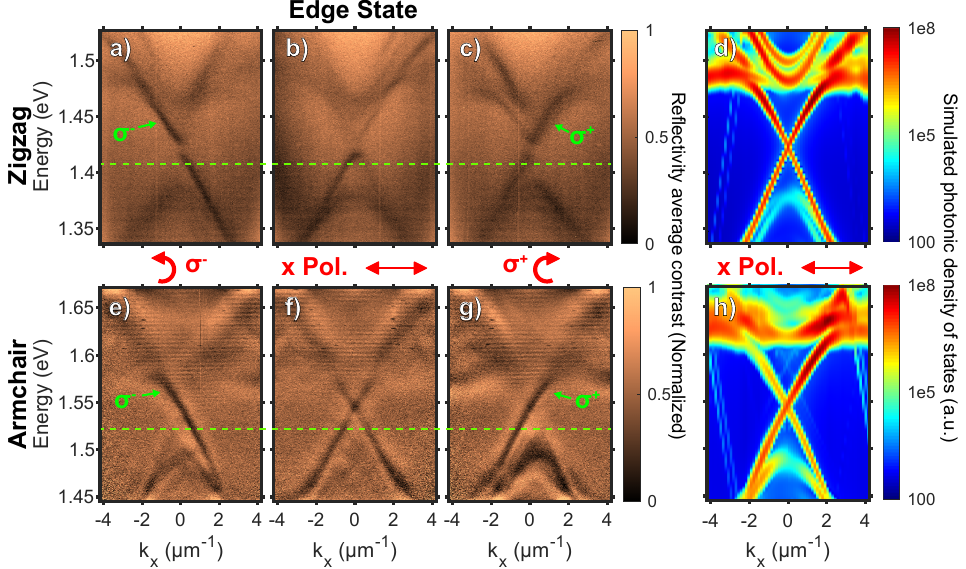}
  \caption{Angle-resolved reflectance contrast data for the interface between the trivial and topological regions for respectively the a) counter-clockwise polarization ($\sigma^-$), b) with polarization along the direction of the interface $x$, and c) clockwise ($\sigma^+$) polarization in the zigzag sample (Fig.~\ref{fgr:fab}) and d) the photonic band structure simulated for linear polarization. e-g) Corresponding reflectance contrast data and h) simulations for the armchair sample. The excitation wavelengths (dashed lines) used in the propagation experiments (Fig. ~\ref{fgr:propexp}) and the unidirectional interface modes (marked $\sigma^-$ and $\sigma^+$) are highlighted in the plots.}
  \label{fgr:edgestate} 
\end{figure*}

Angle-resolved reflectivity contrast measurements shown in Fig.~\ref{fgr:bands} and \ref{fgr:edgestate} were carried out on the \ch{WS_2} zigzag and armchair structures using spatial-filtering within a Fourier-plane spectroscopy set-up (more details in methods and the Supplementary Information), and were compared with finite-difference time-domain (FDTD) simulations (see methods). \\

The trivial (shrunken) and non-trivial (expanded) regions of the two structures were first considered with an incoming linearly polarized white light along the direction of the interface ($x$) (more details in the method section). In the case of the shrunken regions of the zigzag and armchair structures (Fig.~\ref{fgr:bands}a,e), one can observe two parabolic dispersions, one upper mode with a positive effective mass curvature and one lower mode with a negative curvature. For both structures, the upper mode vanishes for low wavevector which is characteristic of a Bound State in the Continuum (BIC) \cite{hsu2016bound} due to the in-plane symmetries of the photonic crystals. The BIC state arises from the destructive interference in the far field of light originating in the photonic crystal. When considering the expanded regions (Figure \ref{fgr:bands}b and f), we can observe that the mode exhibiting a BIC is swapped with the other mode and acquires a negative curvature. This observation is indicative of band inversion and the change of the topological nature of the shrunken and expanded regions. In terms of topology, the bands of the topological (expanded) lattices at the gap are known to have non-vanishing spin-Chern numbers C$_{ph}$ = -1 (upper) and +1 (lower), while the shrunken lattices are topologically trivial with C$_{ph}$ = 0.\cite{li2021experimental}. The dispersions from the shrunken and expanded regions are well reproduced with FDTD simulations in Figure~\ref{fgr:bands}c,d,g and h, with simulated parameters close to the fabricated structure dimensions. In experiments, the expanded lattice can have slightly larger triangles due to a proximity effect in exposure, which can explain the blueshift in bands in \ref{fgr:bands}d. The BIC is not seen in the band structure simulations since they are near-field in character with both the sources and the monitors within the lattices.  \\


We note that due to the change of the triangle size $s$, the optical modes are located at two different spectral regions with mid-gap energies of 1.43 and 1.55 eV, respectively, for the zigzag and armchair structures. This shows the versatility of the studied approach employing van der Waals materials enabling photonic crystal structures for which the spectral position of the modes can be varied by modifying the period, lattice expansion, triangle size, or the thickness of the \ch{WS_2} flake (see Supplementary Figures \ref{fig:bands}, \ref{fig:expansion} and \ref{fig:shape}.\\


The interfaces between the two topologically distinct regions were then studied by collecting reflectivity signal from long and narrow rectangular regions parallel to the interfaces (x-axis in Fig.~\ref{fgr:topo}b, more details in methods). Figure \ref{fgr:edgestate} a) to c) (e) to g)) presents the reflectance of the zigzag (armchair) structures at the interface between the trivial and non-trivial regions for respectively the counter-clockwise ($\sigma^-$), $x$-linear as in in Fig.~\ref{fgr:topo}b, and clockwise ($\sigma^+$) polarizations. In the case of the $x$-linear polarized light (Figure~\ref{fgr:edgestate}b,f), we can observe two modes with linear dispersions with slopes of opposite signs occurring within the gap between the two parabolic photonic mode of the shrunken and expanded regions described previously. These linear modes are the counter-propagating edge states modes occurring at the interface between the trivial and non-trivial regions whose propagating directions are given by the signs of their slopes. To confirm that these modes are indeed edge states, we performed the same reflectivity contrast measurements for the counter-clockwise ($\sigma^-$) and clockwise ($\sigma^+$) polarizations (see Figure~\ref{fgr:edgestate}a,c and e,g). We observed that only the edge-state mode with positive (negative) slope occur for the counter-clockwise (clockwise) polarization. This means that we can select the propagation direction of the edge state mode by switching from counter-clockwise to clockwise polarization. \cite{barik2016two,jalali2020semiconductor}\\

One can notice that in the case of the zigzag structure, a small gap is opened at the intersection between the two edge state modes (see Figure~\ref{fgr:edgestate}b). This gap is also visible when the edge state modes are isolated in the counter-clockwise and clockwise polarization (see Figure~\ref{fgr:edgestate}a,c). This small gap is seen in some cases due to broken crystal symmetry at the interface\cite{barik2016two, xu2016accidental}, and has been attributed to spin-spin scattering in recent studies indicating intrinsic limits of topological protection \cite{parappurath2020direct,wu2015scheme}. \\

\begin{figure*}[!ht]
    \includegraphics[width=\linewidth]{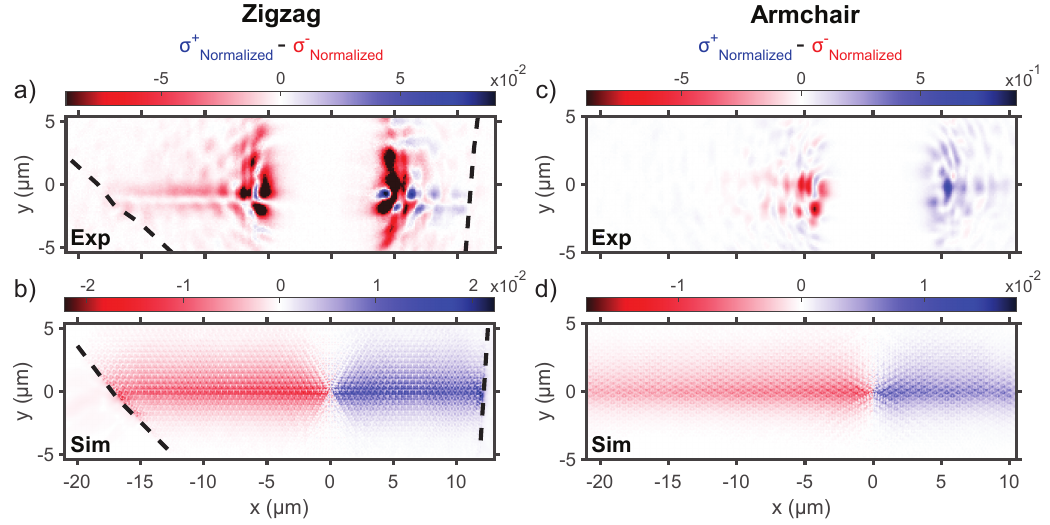}
  \caption{Light propagation along the domain boundary in the spin-Hall WS$_2$ structure. Experimental intensity differences at TE (E field in plane) polarization for a) zigzag and c) armchair structures. The min-max normalized clockwise ($\sigma^+$) signal (normalized between minimum and maximum values) is subtracted from the min-max normalized counter-clockwise ($\sigma^-$) one. b,d) Corresponding differences of the simulated electric field in the zigzag and armchair structures. The black dashed lines in a) and b) indicate the border of the flake of the zigzag structure (see Figure \ref{fgr:fab}a). The white area in the middle of the experimental signals corresponds to the spatial filtering used to suppress the direct intense reflection from the input laser.}
  \label{fgr:propexp}
\end{figure*}

The propagation of the edge states at the topological interface of both structures have been further studied both experimentally and in simulation. The zigzag and armchair topological structures were resonantly excited with an input laser respectively at 881 nm (1.41 eV) and 815 nm (1.52 eV) within the gaps of the band structures (see arrows in Fig.~\ref{fgr:edgestate}). The areas where the propagation was studied  experimentally for both structures are shown Figure \ref{fgr:fab}a and c. The laser was focused in the middle of these areas at the border between the shrunken and expanded regions. The scattered light from the excitation point was then collected with the Fourier spectroscopy set-up and projected to the spectrometer CCD camera. In order to suppress the direct intense reflection from the input laser, the scattering signal was filtered in real space with the use of a wire placed on a lens mount (more details in the method section). Using this technique, long exposure times could be used to observe the emitted light away from the excitation point. The intensity difference between the normalized signals of the clockwise ($\sigma^+$), and counter-clockwise ($\sigma^-$) TE (E field in plane) circular polarizations are obtained in this way and are shown in Figure \ref{fgr:propexp}. The signals for both polarizations are shown in Fig. \ref{fig:propresults} of the SI.  \\


Figure \ref{fgr:propexp} a) and b) show, respectively, the experimental and simulated intensity differences of the zigzag structure with the propagation stopping at flake edges (black dashed lines, see Figure \ref{fgr:fab}a). In the experiment we observe a strong scattering from the laser with a white strip in the middle corresponding to the spatial filtering mentioned previously. However, away from the directly scattered signal, one can observe signal stemming from unidirectional propagating modes, propagating leftward for counter-clockwise polarization (in red), and rightward for the clockwise polarization (in blue) and in both directions for linear polarization. Indeed, as expected, propagation is blocked at the gap for the shrunken and expanded lattices, but is allowed at the interface. 

For the counter-clockwise polarization, the propagation length was estimated to be of 9 µm and 50 µm respectively for the experiment and simulations (see Figures \ref{fig:proplength_sim} and \ref{fig:proplength_exp} of the SI). In the simulation vertical sidewalls are assumed, and inconsistencies in the shape and size of different holes are not accounted for, which would explain for the increased distance in propagation. The selectivity ratio between the polarizations is 23 in simulation, showing good unidirectionality for the interface modes.

Similar results are observed from the armchair structure (see figure \ref{fgr:propexp}c-d), but with much smaller propagation lengths which limits the good observation of the propagation. This difference in the propagation lengths is attributed to the reduced quality of the structure and unoptimally large triangle holes achieved in fabrication for the case of the armchair structure compared to the zigzag structure. Nevertheless, the observation of edge state modes within the gap of the photonic band structure in reflectance measurements, and the direct experimental observation in real space of the selective unidirectional propagation, both well reproduced with FDTD simulations, confirm the topological nature of propagation at the interface of the \ch{WS_2}-based spin-Hall lattice domains. \\


\section{Conclusion}

In conclusion, we have demonstrated the feasibility of using bulk TMD structures for photonic topological insulators with their associated unidirectional interface states. The fabricated insulators have a clear bandgap seen in reflection, and the interface states can be selected with the handedness of circular polarization. The selectiveness in propagation is clearly seen, with a decay length of 9.2 \textmu m measured for the zigzag $\sigma^-$ interface mode compared to 50 \textmu m when the structure is simulated.  This initial demonstration uses a spin-Hall design with modes at $\Gamma$ point above the light cone for ease of measurement and as such the propagation is lossy. Using valley-Hall lattices and separate outcouplers in further experiments will substantially improve these metrics.

Our results emphasize favourable properties of quasi-bulk TMDs, and more generally layered van der Waals materials, and their suitability for fabrication of complex photonic structures. We demonstrate that similarly to silicon-on-insulator structures, and in contrast to GaAs topological photonic crystals, we did not have to rely on suspended membranes for achieving high refractive index contrast necessary for optical confinement. Going forward, the use of van der Waals materials opens new avenues for hybrid topological photonics relying on hetero-integration using pick-and-place nanofabrication approaches. 


\section{Materials and Methods}

\textbf{Simulations.} Finite difference time-domain simulations of the photonic structures were performed with Lumerical. The anisotropic refractive index functions for bulk WS$_2$ used in simulations were obtained with ellipsometry\cite{zotev2023van}. The dimensions of the structures on the sample used for the simulations are detailed on Figure \ref{fgr:fab}. The triangles are truncated in the simulations (see Fig. S3 in the SI) to match the imaged structures with the chamfer values $c$ = 0.16 for the zigzag and $c$ = 0.12 for the armchair sample. For band structure simulations, a 3D model of a doubled hexagonal unit cell was used, since Lumerical requires the repeated Bloch cell to be rectangular. An array of dipoles matched to the unit cells was used with time monitors of the electric field, recording intensity at different frequencies. Simulations with a sweep of different momentum values were done to map out the band structure of the structures, running for a time of 3000 fs. Bands at $\Gamma$ point were simulated with sweeps of various parameters. Edge mode band structure simulations were performed with a supercell containing the interface of topological and trivial lattices (a strip of 9 unit cells of each), using Bloch edge conditions in the direction of the interface and perfectly matched layers in other directions. 

Propagation with chiral polarization was simulated using a perpendicular pair of dipole sources with a 180$^\circ$ phase difference between them. The simulation area was 10 \textmu m by 32 \textmu m, which also included the edges of the flake for the zigzag sample. The model was run for 400 fs in the $\sigma^+$ and 700 fs in the $\sigma^-$ case and the electric field was integrated over that time. 


\textbf{Fabrication.} The WS$_2$ topological structures are fabricated on silicon substrates purchased from Inseto with 1 \textmu m of thermally grown silicon dioxide. First, the substrate is cleaned by acetone (10 min) and isopropanol (IPA) (10 min) in an ultrasonic water bath then blow-drying with nitrogen, after that we treat the substrate in oxygen plasma to remove the residues and contaminants. During the cleaning procedure, WS$_2$ flakes were prepared by the mechanical exfoliation method, repeatedly cleaving with a PVC semiconductor wafer processing tape, on commercially available WS$_2$ crystals (HQ Graphene, synthetic), and then transferred on the clean substrate right away \cite{huang2020universal}. The thickness of the flake was confirmed by atomic force microscopy (AFM Dimension Icon). After we find flakes close to the target thickness (70 nm), we check the candidate flakes by 100x microscope to make sure the surface of the flake is uniform and flat and measure the crystal axis of the flake compared to the bottom edge of the substrate (Fig. \ref{fig:SI_Xuerong} in the SI). 

Second, the sample is prepared for electron beam lithography (EBL). The spin coating procedure consists of two steps, each involving the deposition of the respective film and successive baking on a hotplate. First, spin coating the positive electron-beam resist CSAR 62 (AR-P 6200.13) of thickness 350 nm, which is then covered with a conductive layer of Electra 92 (AR-PC 5090.02) to mitigate charging effects that could reduce the patterning resolution during EBL. In the EBL step, the topological structure design is patterned into the resist layer using an EBL machine (Raith VOYAGER). The pattern for electron beam lithography has exaggerated pointed triangles to improve the resulting shape of the holes in the photonic crystal. The pattern is oriented so that the triangle sides of the pattern coincide with the zigzag crystallographic edge of the individual WS$_2$ flake to take advantage of the anisotropic dry etch in improving the shape of the triangle. To determine the optimum exposure dose, a dose test is typically performed prior to fabrication of the actual sample layout. This entails patterning multiple copies of a test design, each with slightly different exposure doses, which are then checked in a SEM to find the optimum value. 


After the patterning process, the sample is immersed in DI water to remove the layer of Electra 92, then in xylene to dissolve the exposed areas of the positive resist CSAR 62, and finally IPA to get rid of chemical residues prior to blow-drying. The patterned layer of resist covering the sample surface acts as a mask for the subsequent reactive ion etching (RIE), upon which the design is transferred into the WS$_2$ flake. Before etching our sample, we clean the chamber by Ar+H$_2$ and O$_2$ plasma. We use a combination of CHF$_3$ and SF$_6$ plasma to provide a mixture of physical and chemical etching that gives us the best topological structures. After successful RIE, the residual resist film is removed by immersion in hot 1165 resist remover (90$^\circ$C, 30 min) and hot acetone (90$^\circ$C, 30 min), followed by a rinse with IPA and several seconds of O$_2$ ashing to remove the harder resist residues caused by RIE.

\textbf{Optical measurements.} Angle-resolved reflectivity contrast measurements shown in Figs.~\ref{fgr:bands} and \ref{fgr:edgestate} were carried out using a spatial-filtering in a Fourier spectroscopy set-up (see Fig. \ref{fig:dispmeas} in SI). The sample is illuminated in a large region by a collimated white light using a 0.7 NA objective (100X Mitutoyo Plan Apo NIR) and a 150 mm lens before the objective. The reflected light is collected by the same objective, and is separated from the input signal with a beam splitter. The image of the sample is then projected by the objective and a 250 mm lens onto a double slit which selects the desired rectangular region of the sample. A 600 mm lens placed at the focal length behind the spatially-filtered real space performs the Fourier transform of the signal. The Fourier space located at the focal length behind the 600 mm lens is then projected with a set of two lenses (200 mm, 150 mm) onto the slit of a spectrometer which selects the wavevectors along the vertical direction. The diffractive grating inside the spectrometer disperses the light horizontally and the signal is projected onto a 400x100 CCD camera resulting in a ($k_x$,$\lambda$) reflectivity dispersion signal from the rectangular region of interest.

In the case of the measurements of the trivial (shrunken), and non-trivial (expanded) regions, the reflectivity measurements were done using a large rectangular region on the structures (see Fig. \ref{fig:edge_state_step} in the SI). In the case of the measurements at the interface of the two distinct topological regions, the signals were taken from a long and thin rectangular region, parallel to the interface, in order to isolate the signal from the edge states (see Fig. \ref{fig:edge_state_step} in the SI). For each measurement of the structures, another measurement was taken from the unpatterned part of the \ch{WS_2} flake as a reference in order to perform the reflectivity contrast processing. 

The real-space propagation measurements were performed using the same Fourier spectroscopy set-up in which the last two lenses were replaced by a 750 mm lens (see Fig. \ref{fig:propmeas} in the SI). This large focal lens permits to project the real space image with a larger magnification to the CCD camera instead of the Fourier space in the case of the dispersions measurements. The photonic structure were excited with a filtered output from a supercontinuum laser whose wavelengths were put in resonance with the edge states modes. A wire placed on a lens mount was placed in the first real space after the objective (see Fig. S9 in SI), in order to spatially filter the direct reflection of the input laser. Using this technique, long exposure times could be used to observe the scattered light away from the excitation point.  

\textbf{Data processing.} In the case of the reflectivity measurements on the zigzag structure, the reflectivity contrast signal considered was the subtraction of the structure signal by the unpatterned flake signal ($S=S_{structure}-S_{reference}$). However, in the case of the armchair structure, a parasitic Fabry-Perot resonance from the thick \ch{SiO_2} layer hindered the visibility of the optical modes. The signals from the structure and the unpatterned part of the structure were fitted at each $k_x$ slice with a Gaussian line corresponding to the Fabry-Perot mode, which was then removed from the signals. Both the signals were then smoothed in a similar fashion as in ref \cite{li2021experimental}, before performing the same treatment as previously ($S=S^{treated}_{structure}-S^{treated}_{reference}$). More details of the data treatment are given in Supplementary Figures \ref{fig:data_treatment}, \ref{fig:FP_fit} and \ref{fig:avg_contrast}.   



\begin{acknowledgement}
All authors acknowledge the EPSRC grant EP/V026496/1. AIT, DNK and MSS were also supported by the EPSRC grant EP/S030751/1. AIT acknowledges support by the EPSRC grant EP/V006975/1, EP/V007696/1. YW acknowledges a Research Fellowship awarded by the Royal Academy of Engineering RF/201718/17131 and EPSRC grant EP/V047663/1. We thank Luke Brunswick, Dominic Hallett and Luke Wilson for useful discussions.

\end{acknowledgement}





\clearpage
\renewcommand{\thefigure}{S\arabic{figure}}

\setcounter{figure}{0}

\large{\textbf{Supplementary Information: Realization of Z$_2$ topological photonic insulators based on bulk transition metal dichalcogenides}}

\vspace{30pt}

\normalsize{Supplementary Figures S1-S13 with simulated band structures, additional propagation simulations and data, optical measurement methods, data treatment and fabrication details.}

\vspace{30pt}

\begin{figure}[!htb]
\centering
\includegraphics[width=0.8\textwidth]{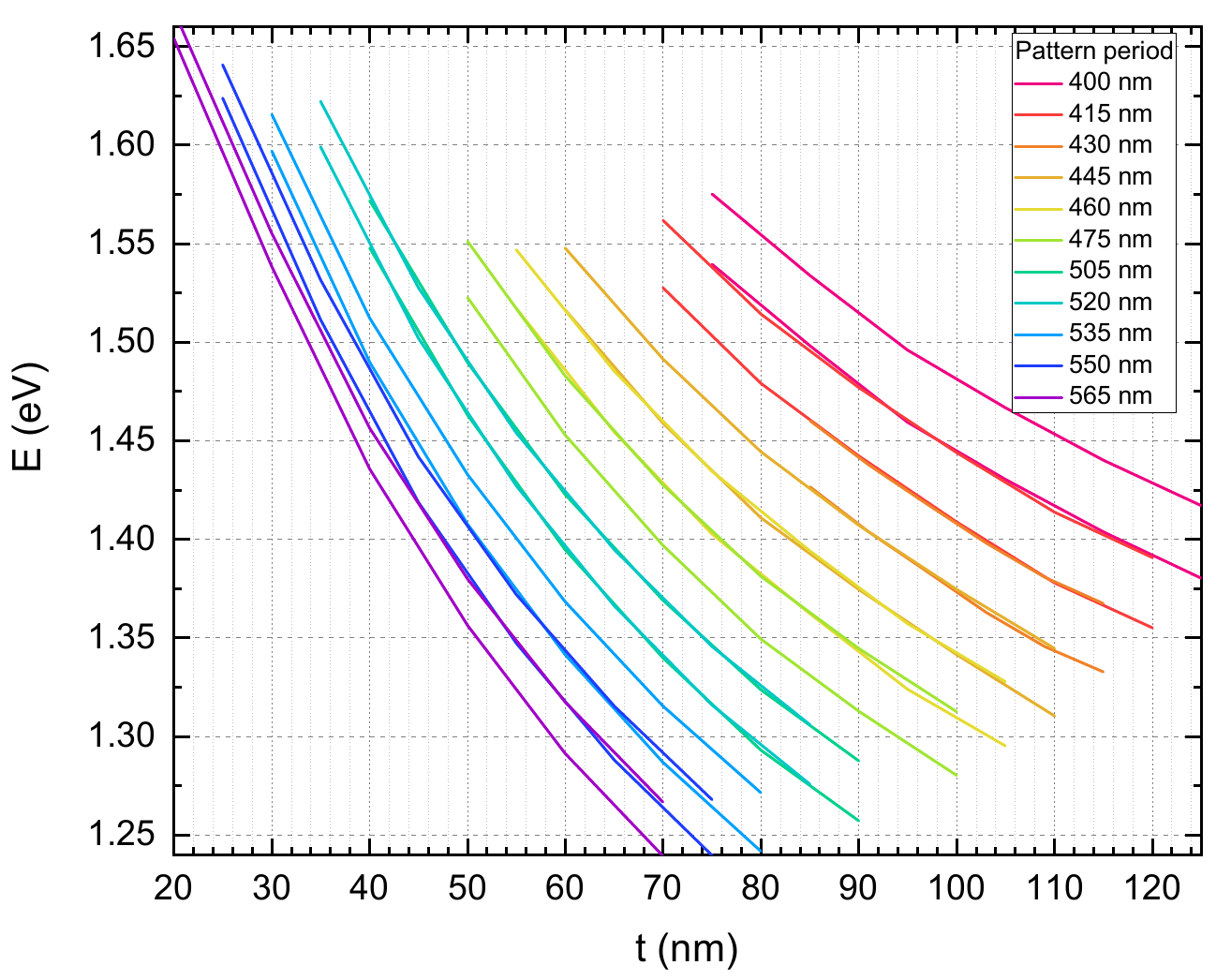}
    \caption{Calculated TE-like photonic band gaps at $\Gamma$ of a shrunken (0.95) hole lattice of triangles with their side length kept at $s = 0.315a$. The thickness $t$ of the WS$_2$ flake is varied on a 1 \textmu m SiO$_2$/Si substrate. Lines with different colours correspond to different periods $a$.}
\label{fig:bands}
\end{figure}

\begin{figure}[!htb]
\centering
\includegraphics[width=0.6\textwidth]{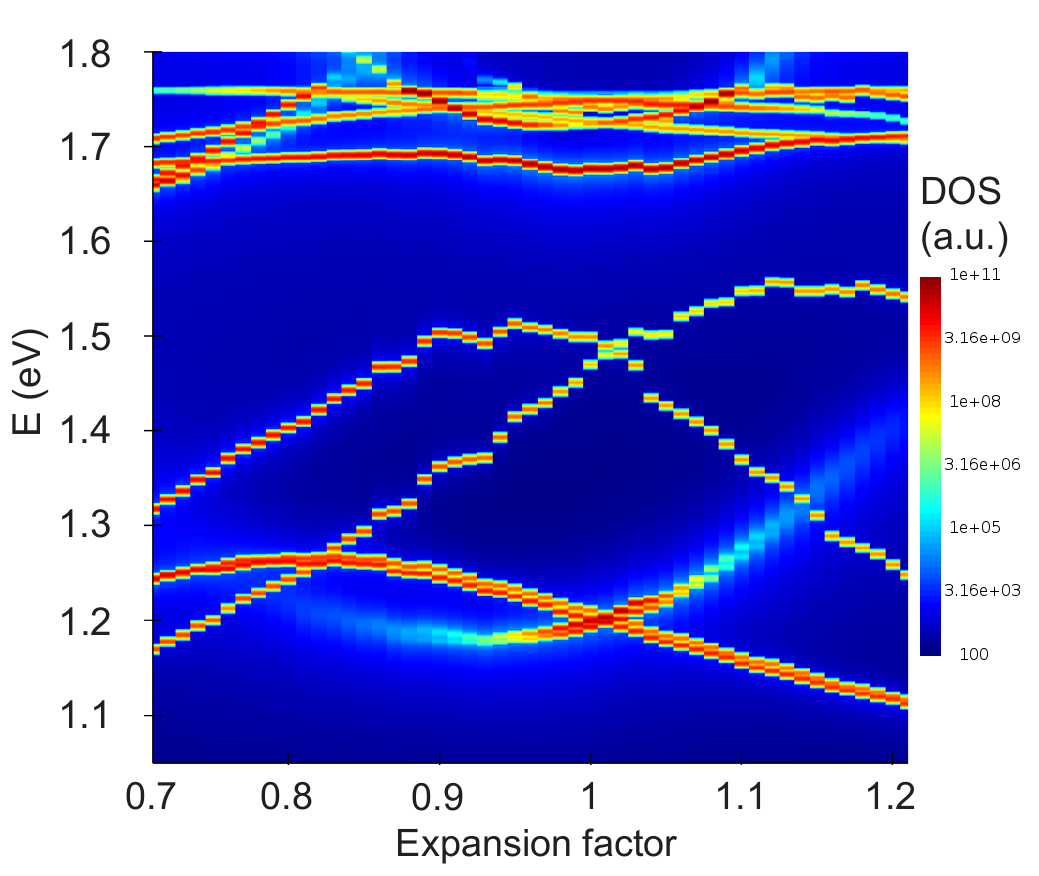}
    \caption{Calculated TE-like photonic bands of spin-Hall lattices in WS$_2$ at $\Gamma$ with various expansion factors. The relevant bands cross at 1.2 eV at factor 1. $a$ = 560 nm, $h$ = 80 nm, $c$ = 0.}
\label{fig:expansion}
\end{figure}

\begin{figure}[!htb]
\centering
\includegraphics[width=0.9\textwidth]{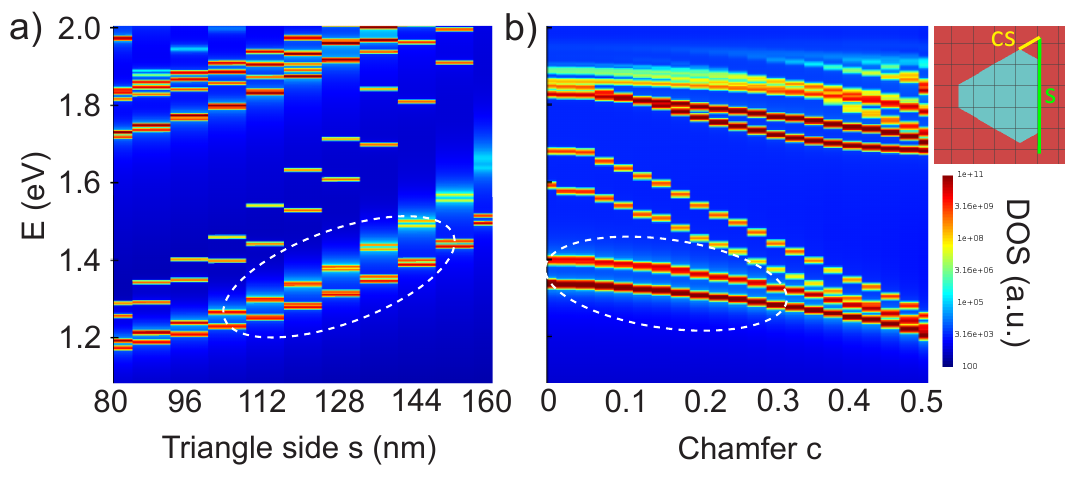}
    \caption{Calculated TE-like photonic bands at $\Gamma$. $a$ = 440 nm, $h$ = 100 nm, expansion factor 1.05. a) Triangle size (with $c$ = 0) is changed. The energy of the bands increases as the triangle holes increase in size and the effective refractive index of the film decreases. b) Cutoff of the triangle corners (with $s$ = 138 nm) is varied. In both cases the relevant gap is highlighted. The chamfer parameter is defined with the distance cut from every corner of each triangle. Illustration of a triangle hole with a chamfer value $c$ = 0.2 is shown in the inset.}
\label{fig:shape}
\end{figure}




\begin{figure}[!htb]
\centering
\includegraphics[width=\textwidth]{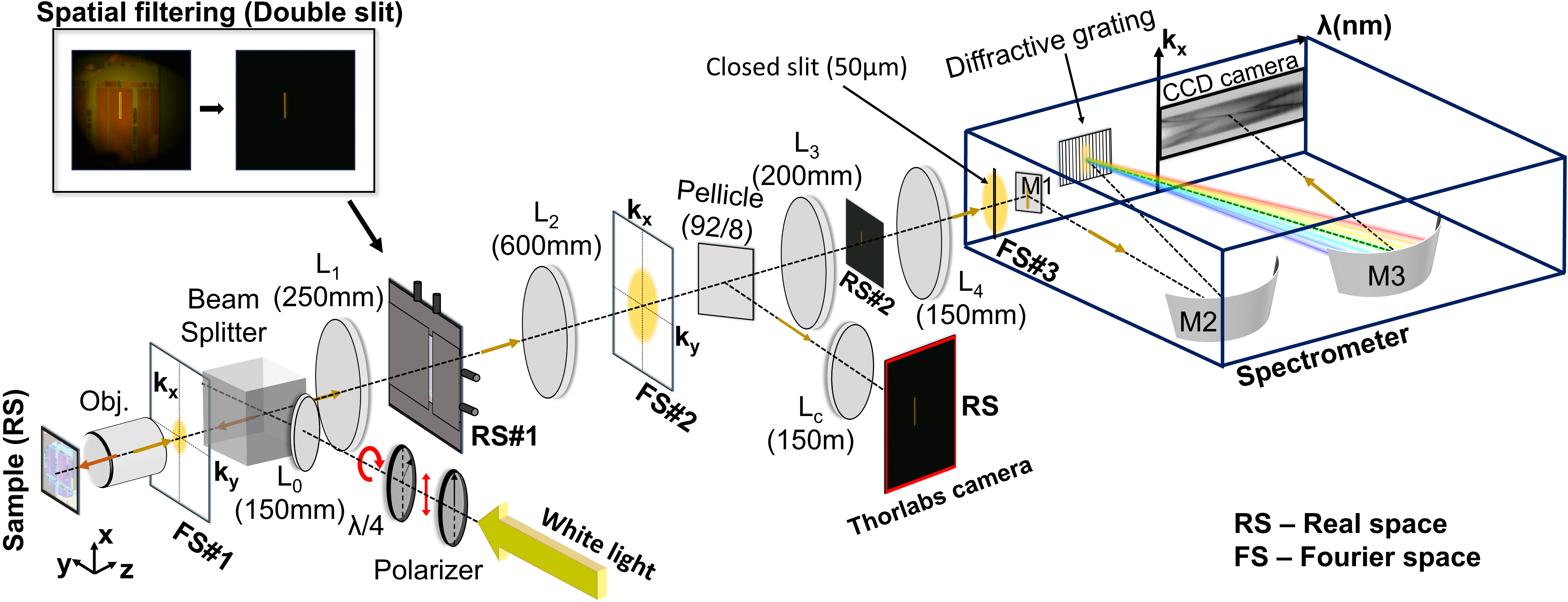}
    \caption{Angle-resolved reflectivity contrast measurements were carried out using a home-made spatial-filtering Fourier set-up. The sample is illuminated in a large region by a collimated white light using a 0.7 NA objective (100X Mitutoyo Plan Apo NIR) and a 150 mm lens ($L_0$) before the objective. The reflected light is collected by the same objective, and is separated from the input signal with a beam splitter. The image of the sample is then projected by the objective and a 250 mm lens ($L_1$) onto a double slit which selects the desired rectangular region of the sample. A 600 mm lens ($L_2$) placed at focal length behind the spatially-filtered real space performs the Fourier transform of the signal. The Fourier space (\#2) located at focal length behind the 600 mm lens is then projected with a set of two lenses (200 mm ($L_3$), 150 mm ($L_4$)) onto the slit of a spectrometer which selects the wavevectors along the vertical direction. The diffractive grating inside the spectrometer disperses the light horizontally and the signal is projected onto a 1340x400 CCD camera resulting in a ($k_x$,$\lambda$) reflectivity dispersion signal from the rectangular region of interest.}
\label{fig:dispmeas}
\end{figure}

\begin{figure}[!htb]
\centering
\includegraphics[width=\textwidth]{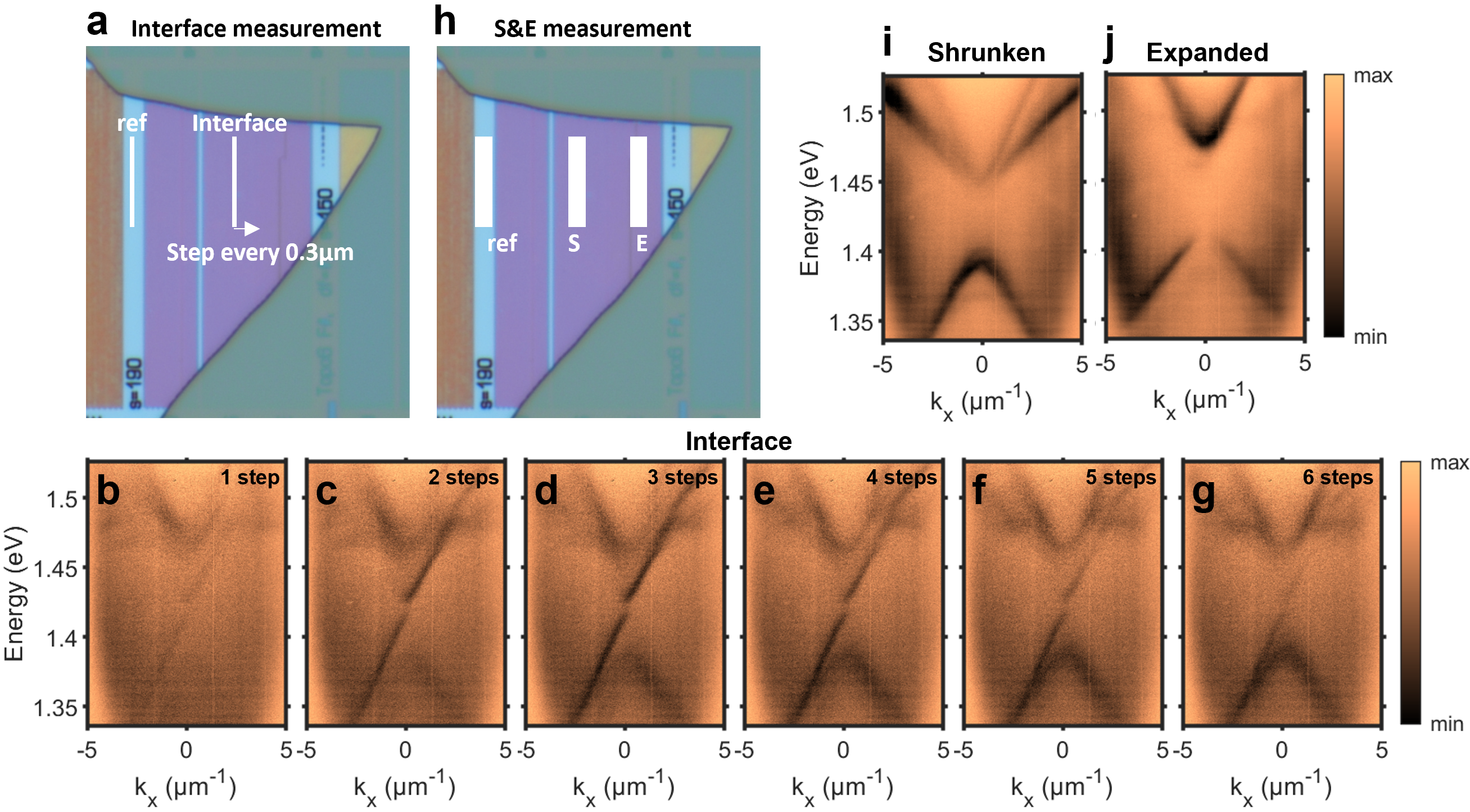}
    \caption{Method of the band structure measurements of the interface and the shrunken and expanded lattices. (a) Interface measurement method. An 1x20 \textmu m$^2$ rectangular region is selected with the double slit placed in the real space (see figure \ref{fig:dispmeas}). The reflectivity measurements were first performed on the unpatterned part of the flake as reference signal, and several ones taken 16.8\textmu m + nx0.3\textmu m steps on the right. The resulting results at different steps are shown in (b) to (g). This method was used to obtain the best region of the interface modes. (h) Shrunken and expanded lattices measurement method. A bigger rectangular region is considered (2.5x20 \textmu m$^2$), and the resulting results are shown in (i) and (j). The treatment method of this measurement are further detailed in figures \ref{fig:data_treatment}, \ref{fig:FP_fit}, and \ref{fig:avg_contrast}. 
}
\label{fig:edge_state_step}
\end{figure}

\begin{figure}[!htb]
\centering
\includegraphics[width=\textwidth]{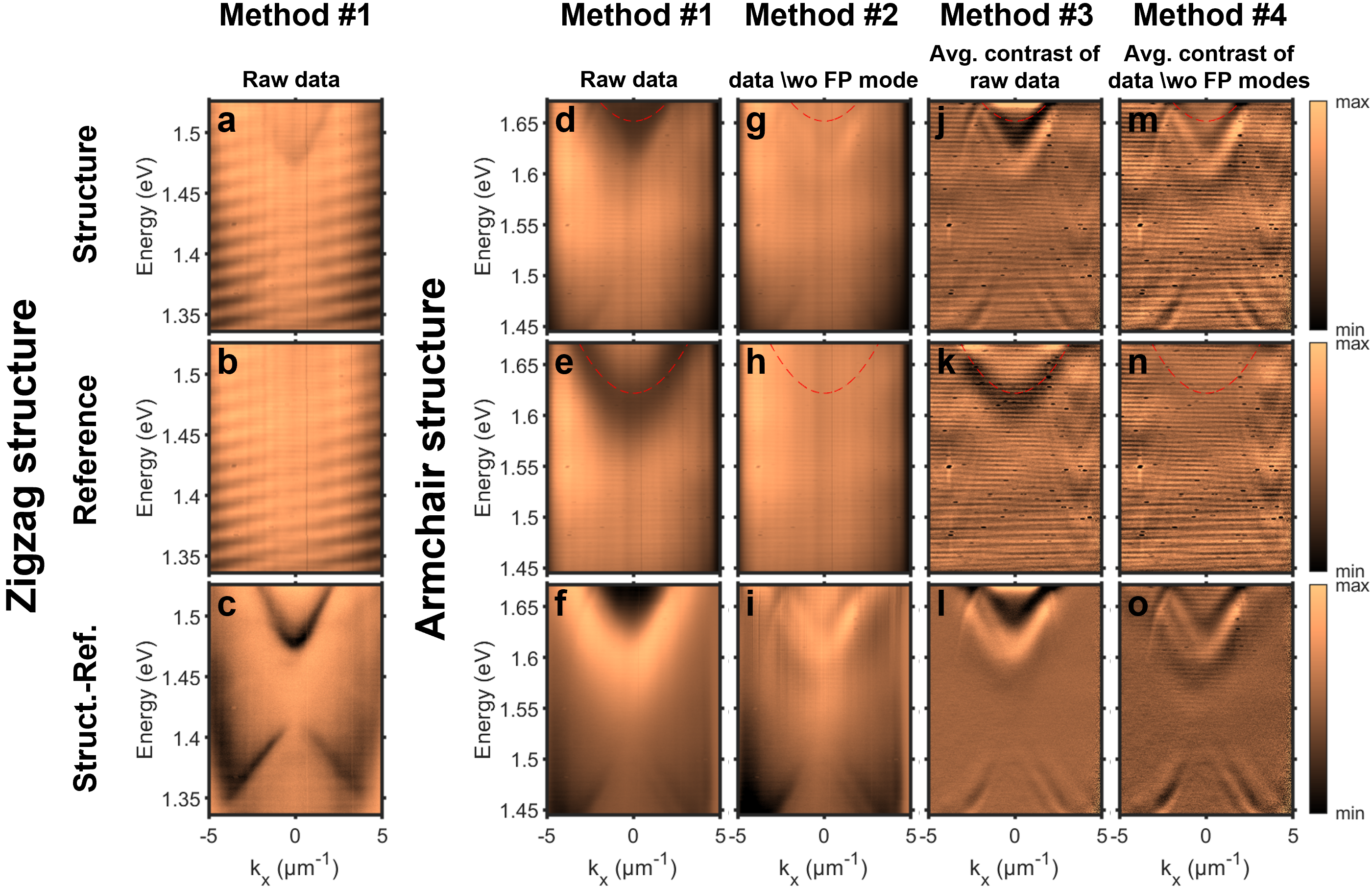}
    \caption{Data treatment of the band structure reflectivity measurements. a) to c) reflectivity measurement of the zigzag structure expanded region with (a) the structure reflectivity measurement, (b) the flake reference measurement and (c) reflectivity difference between the structure and reference measurement. d) to f) reflectivity measurement of the armchair structure expanded region. In this case, a Fabry-Perot (FP) mode is localized directly on top of the expanded top mode. Moreover, the FP mode is redshifted between the structure (d) and reference (e) measurements as the two regions possess slightly different effective refractive index. When plotting the reflectivity difference (f), the two shifted FP modes are visible, which hide the expanded upper mode. g) to i) Structure (g) and reference (h) signals where the FP modes were fitted and removed (see method in Figure \ref{fig:FP_fit}). The upper mode of the photonic band structure is visible in differential reflectivity signal (i), but with low contrast. j) to l) Structure (j) and reference (k) signals smoothed in a similar fashion as in \cite{li2021experimental} (see method in Figure \ref{fig:avg_contrast}). The differential reflectivity (l) possess a better contrast but the upper mode is still hidden by the FP modes. m) to o) Structure (m) and reference (n) signals for which the FP modes were removed and then smoothed. The differential reflectivity (o) possess a good contrast and upper mode visibility using the combination of the FP mode removal technique and smoothing.
}
\label{fig:data_treatment}
\end{figure}

\begin{figure}[!htb]
\centering
\includegraphics[width=\textwidth]{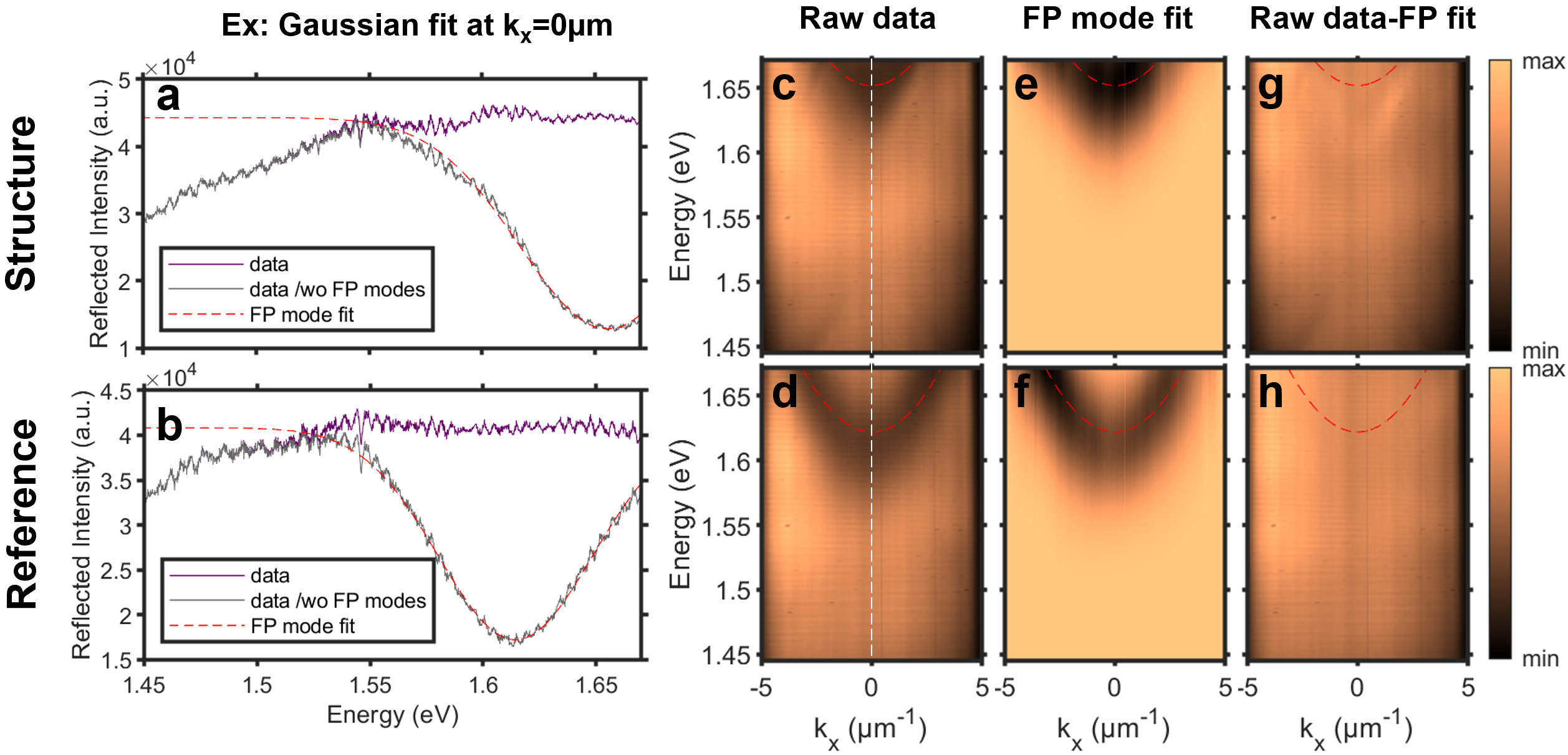}
    \caption{Fabry-Perot (FP) mode fitting method for the armchair structure measurements. Each vertical slices at given $k_x$ of the reference (c) and structure (d) raw data measurement were fitted with a Gaussian line to remove the Gaussian FP resonance. Examples of these fits for $k_x$ = 0 are shown for the reference (a) and structure (b) signals. The resulting FP mode fitting for all the vertical slices are shown in the colormaps in (e) and (f). The difference between the raw data and the fitted FP modes are shown respectively in (g) and (h) for the reference and structure raw data.
}
\label{fig:FP_fit}
\end{figure}

\begin{figure}[!htb]
\centering
\includegraphics[width=\textwidth]{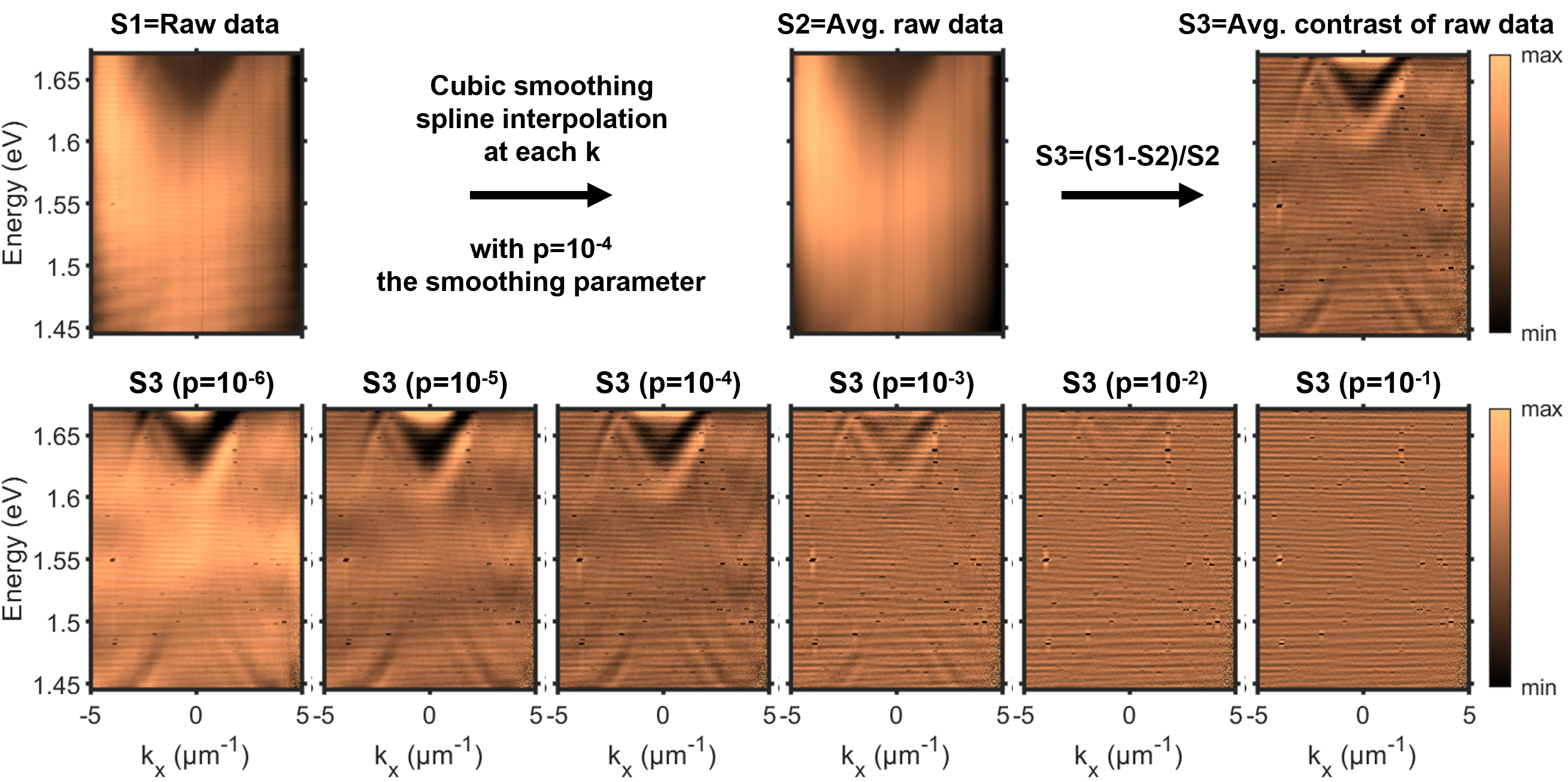}
    \caption{Average contrast treatment method. The raw data signals, S1 (a), is smoothed, S2 (b), using the cubic smoothing spline interpolation Matlab function with a smoothing parameter $p=10^{-4}$. The reflectivity contrast S3=(S1-S2)/S2 is shown in (c). (d) to (i) Average contrast treatment method, for smoothing parameters ranging from $10^{-6}$ to $10^{-1}$. When the smoothing parameter p is too small, $p<10^{-4}$, the Fabry-Perot (FP) mode is not entirely removed. When the smoothing parameter is too high, $p>10^{-4}$, the FP mode is well removed but also the structure optical bands. For the armchair structure it was found that the parameter $p=10^{-4}$ combined with a FP mode removal detailed in Figure \ref{fig:FP_fit} was the optimal method to remove  the FP mode as much as possible while keeping the structure optical modes.
}
\label{fig:avg_contrast}
\end{figure}


\begin{figure}[!htb]
\centering
\includegraphics[width=\textwidth]{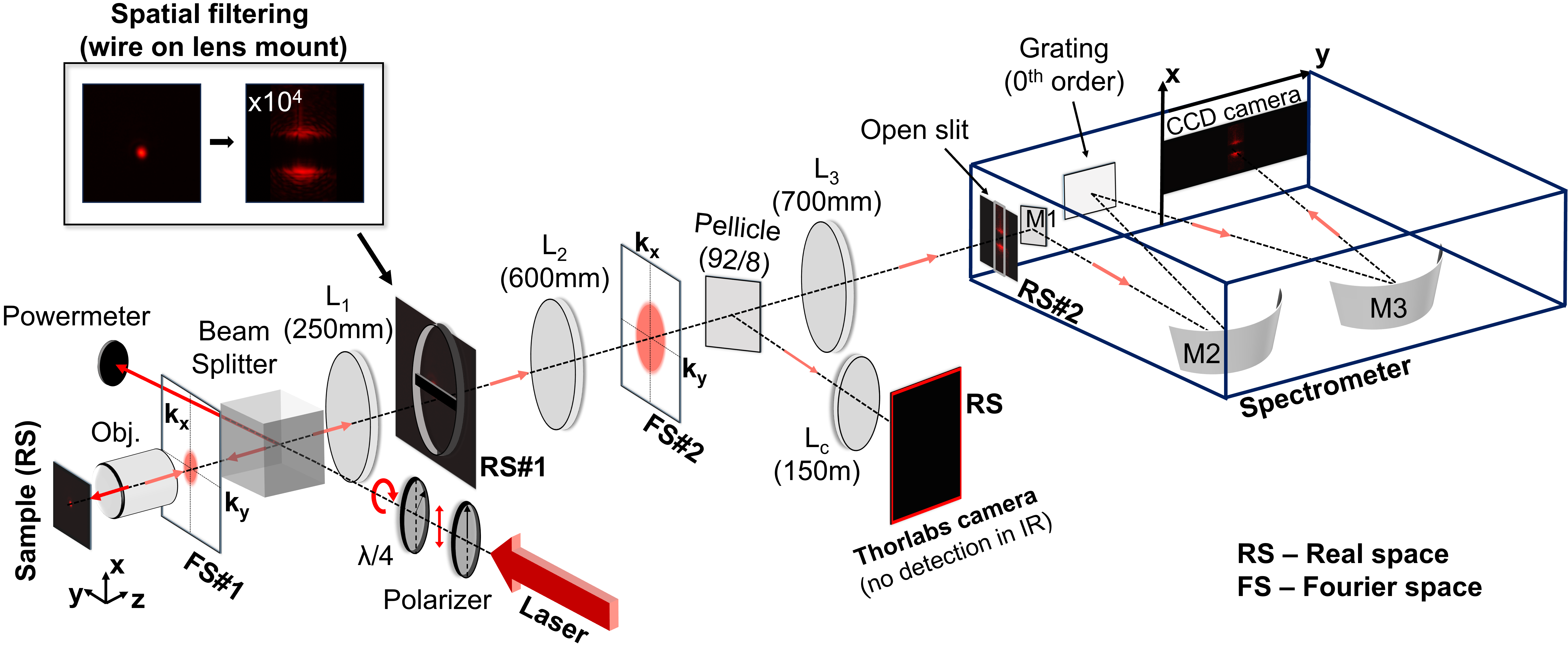}
    \caption{Propagation measurements were carried out using a slightly altered Fourier setup where the Fourier lens L3 and spectrometer lens L4 were flipped down and replaced by a spectrometer lens L5 of large 700 mm focal length in order to get a large real space image onto the CCD camera. The sample image (real space) is projected a first time with the 100x objective and the 250 mm lens. We proceed to a spatial filtering by placing a wire on the first projected real space. This is to get rid of the intense laser signal at the excitation spot in order to see the propagation signal away from the laser spot. Then, the filtered real space is projected to the spectrometer and goes through the spectrometer slit and is projected onto the CCD camera as the spectrometer grating is put at 0th order and acts as a mirror.
}
\label{fig:propmeas}
\end{figure}

\begin{figure}[!htb]
\centering
\includegraphics[width=\textwidth]{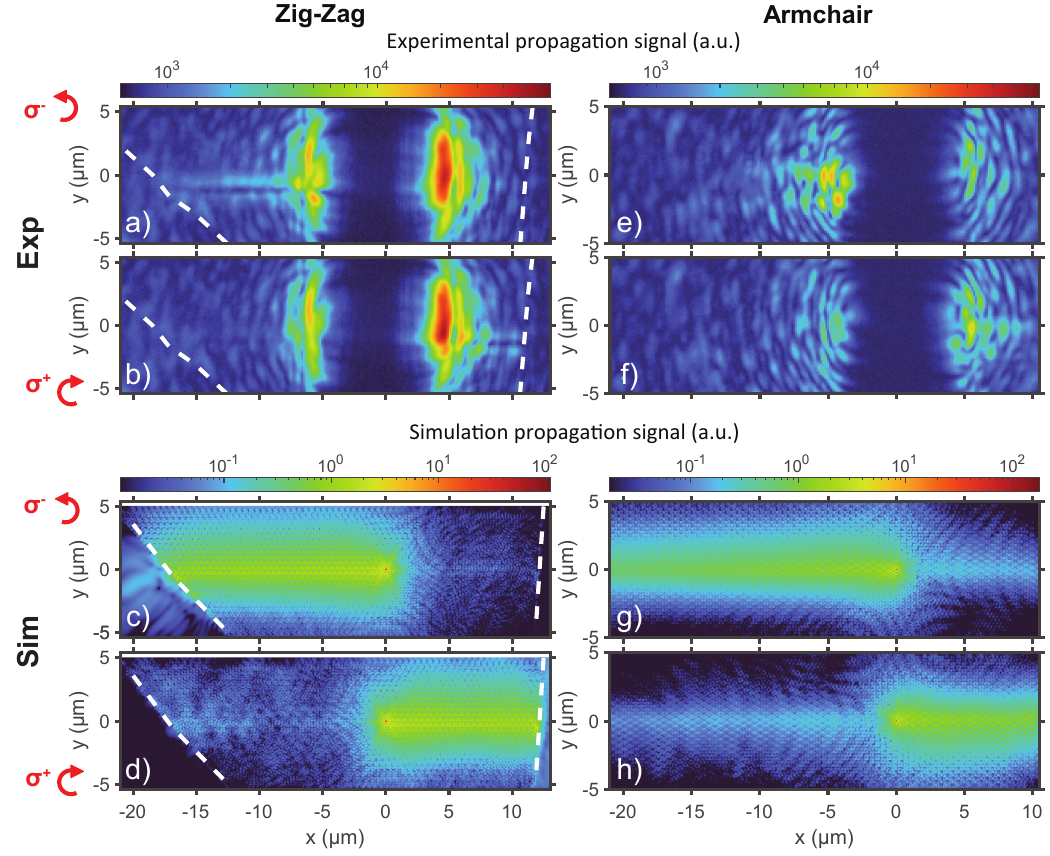}
    \caption{Scattered light coming from the zigzag and armchair structures using the propagation setup in \ref{fig:propmeas}. a) and b) (e) and f)) scattered signal from the zigzag (armchair) structure for respectively the counter-clockwise circular and clockwise circular TE (E field in plane) polarizations. The corresponding FDTD simulations with integrated E field at the center plane of the WS$_2$ layer are shown in c) and d) (g) and h)). The white dashed lines in a-d) indicate the border of the flake in the case of the zigzag structure (see Fig. 2a) in the article). The data used is the same as in Fig. 5 of the article.}
\label{fig:propresults}
\end{figure}

\begin{figure}[!htb]
\centering
\includegraphics[width=\textwidth]{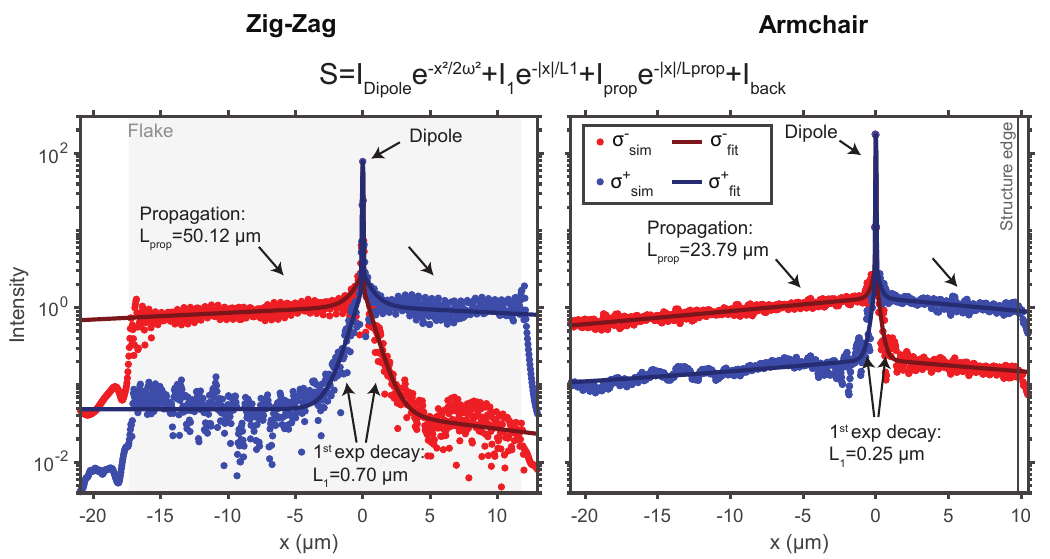}
    \caption{Propagation lengths from the simulation of the zigzag and armchair structures. The cross-sections taken from the simulations shown in figs \ref{fig:propresults} c-d) and g-h) were taken at y=0 \textmu m. The cross-sections of the counter-clockwise polarization simulations ($\sigma^-$, Fig. \ref{fig:propresults} a)) are plotted as red dots, and the clockwise polarization ($\sigma^+$, Fig. \ref{fig:propresults} b)) as blue dots. Both cross-sections were fitted separately for positive and negative x, with the following equation: $S=I_{Dipole}e^{-x^2/2\omega^2}+I_{1}e^{-|x|/L_1}+I_{prop}e^{-|x|/L_{prop}}+I_{back}$, with $I_{Laser}$, $I_{prop}$, $I_1$, $I_{back}$, the dipole, first decay, propagating mode, and background intensities, $\omega$ the dipole waist, and $L_{prop}$ the propagation length. Propagating lengths of 50.12 \textmu m and 23.79 \textmu m were found for respectively the zig-zag and armchair structures. One can notice that there is still propagation along the interface for $\sigma^+$ (blue) polarization for negative x, and $\sigma^-$ (red) polarization for negative x. However, the intensity of these propagation are an order of magnitude lower than for the opposite circular polarization. The directional selection power, corresponding to the signal ratio between the propagation in one direction of the two circular polarizations, $I^{\sigma^-}_{prop(left)}/I^{\sigma^+}_{prop(left)}$ or $I^{\sigma^+}_{prop(right)}/I^{\sigma^-}_{prop(right)}$, are 23.34 and 8.02 for the zigzag and armchair structures, respectively.
}
\label{fig:proplength_sim}
\end{figure}

\begin{figure}[!htb]
\centering
\includegraphics[width=\textwidth]{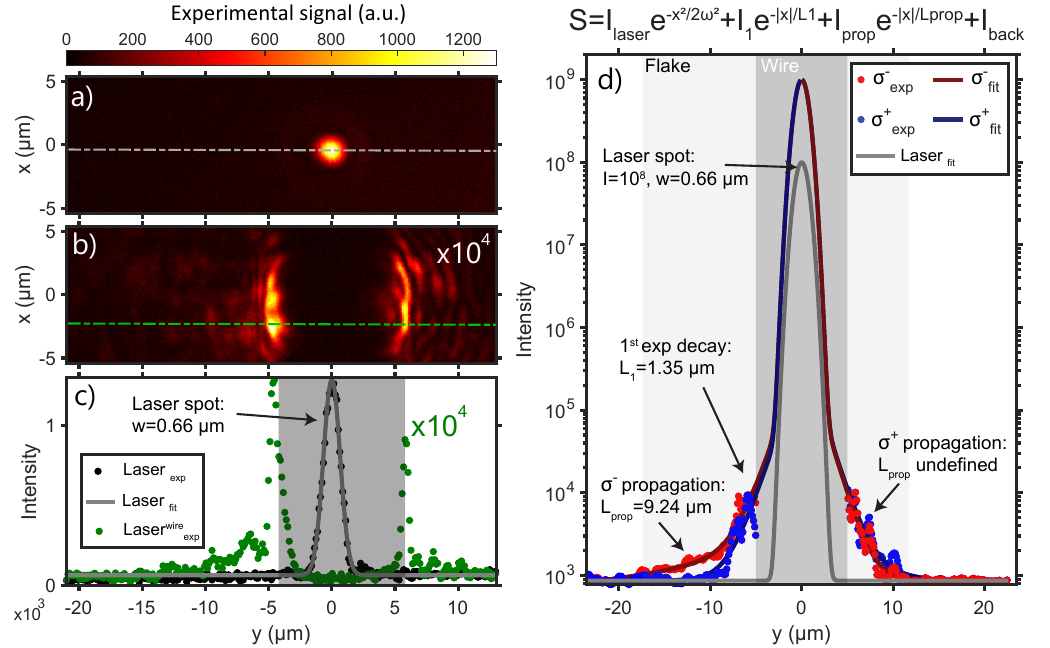}
    \caption{Propagation length estimation from cross-sections of the zigzag structure propagation measurements. a-c) Estimation of the laser waist and the signal reduction when placing a wire as a spatial filter (see Fig. \ref{fig:propmeas}). a) Laser spot signal on the structure away from the interface with ND filters reducing the signal by $10^4$. b) Laser signal on the same region with the wire as a spatial filter, and without the ND filters. Consequently, the signal collected at the border of the wire is $10^4$ weaker than the laser spot. c) Cross sections of the two previous measurements in a-b) indicated by the dashed gray and green lines. The laser spot was fitted with a Gaussian line $S=I_{Laser}e^{-x^2/2\omega^2}$, giving a laser spot waist of 0.66 \textmu m. d) Cross-sections taken from the measurements shown in figs \ref{fig:propresults} a-b) in a range of y in which the interface propagation signals is highest. The cross-section of the counter-clockwise polarization measurements ($\sigma^-$, Fig. \ref{fig:propresults} a)) is plotted as red dots, and the clockwise polarization ($\sigma^+$, Fig. \ref{fig:propresults} b)) as blue dots. The cross-sections were fitted the same way as for the simulation (see Fig. \ref{fig:proplength_sim}) with $S=I_{Laser}e^{-x^2/2\omega^2}+I_{1}e^{-|x|/L_1}+I_{prop}e^{-|x|/L_{prop}}+I_{back}$. The laser waist of 0.66 \textmu m obtained in c), and a laser intensity of $10^8$ ($10^4\times max(signal_{exp})$) were used as fixed parameters. In the case of the $\sigma^+$ (blue) signal for positive x, the poor contrast between the propagating mode signal and laser spot signal prevents from obtaining a good fit and a good value of propagation length. However a very good match is obtained between the experimental $\sigma^-$ (red) signal for negative x and the model, which gives a propagation length of 9.24 \textmu m. }
\label{fig:proplength_exp}
\end{figure}

\begin{figure}[!htb]
\centering
\includegraphics[width=\textwidth]{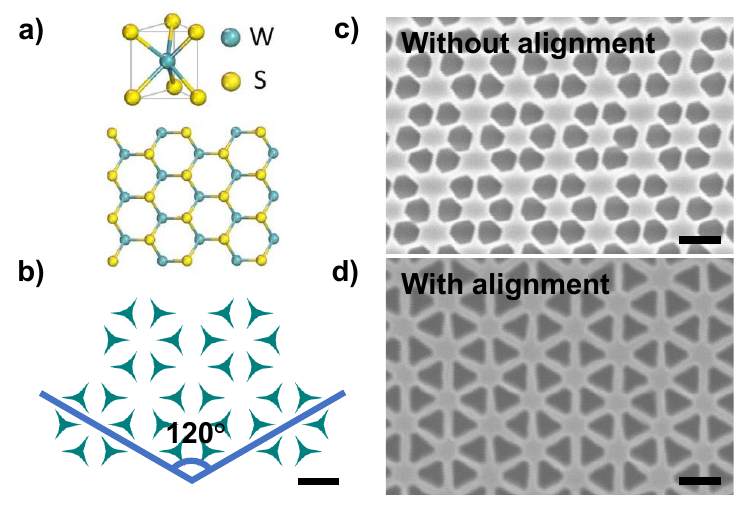}
    \caption{a) Hexagonal crystal structure of WS$_2$. b) The exposed dose pattern with five shrunken lattice elements displayed using an optimized pointed triangle design. The alignment of the pattern to the crystal axis of the WS$_2$ flake is shown, with the pattern aligned to a visible corner (blue lines) on the edge of the flake. SEM images of a c) fabricated structure without alignment or pointed triangle design and an d) optimized structure using flake alignment. Nominal triangle side length $s$ = 150 nm, $a$ = 478 nm, scale bars 200 nm.
}
\label{fig:SI_Xuerong}
\end{figure}

\clearpage


\bibliography{main}

\end{document}